\documentclass[runningheads]{llncs}

\usepackage{graphicx}
\usepackage{tikz}
\usepackage{amsmath, amssymb}
\usepackage[misc]{ifsym} 
\usepackage{bbding}

\usepackage{wrapfig}
\DeclareMathOperator*{\argmin}{\arg\!\min}

\makeatletter
\def\blfootnote{\xdef\@thefnmark{}\@footnotetext}
\makeatother

\usepackage[acronyms]{glossaries}

\newacronym{GAN}{GAN}{generative adversarial network}
\newacronym{CNN}{CNN}{convolutional neural network}
\newacronym{COPD}{COPD}{chronic obstructive pulmonary disease}
\newacronym{PFT}{PFT}{pulmonary function test}
\newacronym{ROI}{ROI}{region of interest}
\newacronym{IRB}{IRB}{Institutional Review Board}
\newacronym{PHI}{PHI}{Protected Health Information}
\newacronym{DeFI}{DeFI-GAN}{deformation field interpretation with generative adversarial networks}
\newacronym{VA-GAN}{VA-GAN}{visual attribution generative adversarial network}
\newacronym{VR-GAN}{VR-GAN}{visualization for regression with a generative adversarial network}
\newacronym{AD}{AD}{Alzheimer's disease}
\newacronym{MCI}{MCI}{mild cognitive impairment}
\newacronym{bet}{BET}{the Brain Extraction Tool}
\newacronym{fsl}{FSL}{FMRIB Software Library}
\newacronym{fov}{FOV}{field of view}
\newacronym{ants}{ANTs}{Advanced Normalization Tools}
\newacronym{mni}{MNI}{Montreal Neurological Institute}
\newacronym{ncc}{NCC}{normalized cross-correlation}
\newacronym{MRI}{MRI}{magnetic resonance imaging}
\newacronym{BMI}{BMI}{body mass index}
\newacronym{ADNI}{ADNI}{Alzheimer's Disease Neuroimaging Initiative}
\newacronym{PA}{PA}{posterioranterior}
\newacronym{WGAN}{WGAN}{Wasserstein generative adversarial network}
\newacronym{CAM}{CAM}{class activation map}
\newacronym{LRP}{LRP}{layer-wise relevance propagation}
\newacronym{IG}{IG}{integrated gradients}
\newacronym{CXR}{CXR}{chest x-ray}

\usepackage{todonotes}
\usepackage[title]{appendix}
\usepackage{booktabs}
\usepackage{bm}
\usepackage{nicefrac}
\usepackage{float}
\usepackage{adjustbox}
\usepackage{xr}
\usepackage{relsize}

\newcommand\bsfrac[2]{%
\scalebox{-1}[1]{\nicefrac{\scalebox{-1}[1]{#1}}{\scalebox{-1}[1]{#2}}}%
}

\begin{document}
\title{Interpretation of Disease Evidence for Medical Images Using Adversarial Deformation Fields}
\titlerunning{Interpretation of Disease Evidence Using Adversarial Deformation Fields}
\author{Ricardo Bigolin Lanfredi\inst{1}\orcidID{0000-0001-8740-5796} \and
Joyce D. Schroeder\inst{2}\orcidID{0000-0002-7451-4886} \and
Clement Vachet\inst{1}\orcidID{0000-0002-8771-1803} \and
Tolga Tasdizen\inst{1}\orcidID{0000-0001-6574-0366}}
% index{Bigolin Lanfredi, Ricardo}
% index{Schroeder, Joyce}
% index{Vachet, Clement}
% index{Tasdizen, Tolga}
\authorrunning{R. Bigolin Lanfredi et al.}
\institute{Scientific Computing and Imaging Institute,\\University of Utah, Salt Lake City UT 84112, USA\\
\email{ricbl@sci.utah.edu} \and
Department of Radiology and Imaging Sciences,\\University of Utah, Salt Lake City UT 84112, USA}
\maketitle
\begin{abstract}
The high complexity of deep learning models is associated with the difficulty of explaining what evidence they recognize as correlating with specific disease labels. This information is critical for building trust in models and finding their biases. Until now, automated deep learning visualization solutions have identified regions of images used by classifiers, but these solutions are too coarse, too noisy, or have a limited representation of the way images can change.  We propose a novel method for formulating and presenting spatial explanations of disease evidence, called \gls{DeFI}. An adversarially trained generator produces deformation fields that modify images of diseased patients to resemble images of healthy patients. We validate the method studying \gls{COPD} evidence in \glspl{CXR} and \gls{AD} evidence in brain MRIs. When extracting disease evidence in longitudinal data, we show compelling results against a baseline producing difference maps. \gls{DeFI} also highlights disease biomarkers not found by previous methods and potential biases that may help in investigations of the dataset and of the adopted learning methods.

\keywords{ Deformation Field \and Deep Learning Interpretation \and Visual Attribution \and Adversarial Training \and Disease Effect \and \gls{DeFI}}

\end{abstract}
\glsresetall

\section{Introduction} \protect\blfootnote{Data used in preparation of this article were obtained from the \gls{ADNI} database (\url{adni.loni.usc.edu}).}

The recent surge of deep learning applications has the potential to revolutionize medical imaging in several ways, such as accessibility, efficiency, and flexibility. However, decision automation leads to challenges, including the interpretability of the outcomes~\cite{roadmap}. Understanding what kinds of evidence deep learning methods capture from an image is one approach for overcoming these challenges. The field can benefit from such understanding through improvements in user trustworthiness, patient communication, and model bias identification.

Among visual attribution methods, backpropagation through a trained classifier is commonly used to determine areas of stronger influence on the model output~\cite{visreview}. Several visual attribution approaches have been applied to medical imaging. The \gls{CAM}~\cite{cam} method has been employed to show low-resolution areas of focus when performing pneumonia diagnosis from a \gls{CXR}~\cite{chexnet}. \gls{CAM} has also been used as part of a  weakly-supervised fine segmentation of lung nodules~\cite{reviewer_added}. \Gls{LRP}~\cite{lrp} has been used to find biases in a histopathology dataset~\cite{biasfound}. Despite their usefulness, most visual attribution methods are too coarse or too noisy for some applications~\cite{vagan}. They also have limited means of representing their findings, which may obscure relevant evidence. 

Spatial deformations have been used in several applications, including registration~\cite{stlregistration}, generation of adversarial examples~\cite{spatialadv}, and creation of atlases~\cite{NIPS2019_8368}. While deformation fields may not be adequate for producing color and texture changes, they can generate variations in position, shape, and size. Since some impacts of diseases on anatomy are linked to the latter, we hypothesize that applying deformation fields to represent and visualize model interpretations will better include such variations. We propose the \gls{DeFI} method, which uses adversarial training~\cite{gan} to learn to produce a deformation field that alters an image to make disease signs indiscernible. Therefore, the changes caused by this field express evidence of such disease. To the best of our knowledge, no other work has used deformation fields for visual attribution in deep learning.

We perform experiments studying evidence of \gls{COPD} in \glspl{CXR}~\cite{emphysema1,emphysema2}. In~\cite{vrgan}, automatic assessment of \gls{COPD} evidence is also performed. However, this assessment applies additive perturbations instead of deformation fields, and a different set of disease evidence is found. The method presented in \cite{vrgan} also models disease severity, which is not a focus of \gls{DeFI}. We also employ the \gls{ADNI} dataset to model the conversion from \gls{MCI} to \gls{AD} through the morphological brain changes observed in MRIs~\cite{brain_main}.  In~\cite{vagan}, generative adversarial training is also used to assess evidence of \gls{AD} in brain MRIs. This setup inspires our method, but uses additive perturbations instead of deformation fields and a different regularization loss function. We analyze the outcomes of the proposed formulation in both datasets, showing that the use of \gls{DeFI} improves longitudinal prediction over a baseline~\cite{vagan} and highlights additional disease evidence.
\section{Method}
 \begin{figure}[t]
 \centering
\includegraphics[width=1.0\textwidth,trim={5.1cm 4.2cm 1cm 15.6cm},clip]{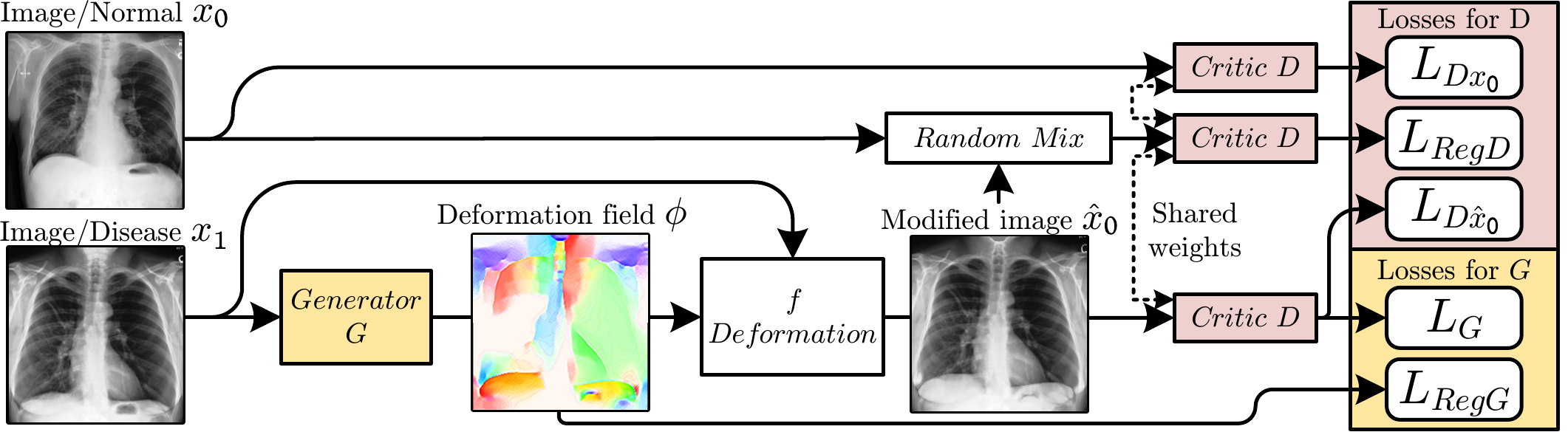}
\caption{Overall model architecture. The terms $L_{Dx_0}$, $L_{D\hat{x}_0}$, and $L_{G}$ are \gls{WGAN} losses, whereas $L_{RegD}$ penalizes the gradient of $D$, and $L_{RegG}$ penalizes the complexity of $\phi$.} \label{diagram}
\end{figure}
We consider binary classification problems, where class 0 is associated with healthy patients and class 1 with diseased patients. Each image in a dataset is denoted by $x$, and its domain by $X$. Images of class $c$ are indicated by $x_c$. Our objective is to find a transformation $f$ that maps from an image $x_1$ containing evidence for a disease to a modified image $\hat{x}_0$ where that evidence is absent. The rest of the content, e.g. patient specific anatomy, should not be modified. We propose to model $f$ as a deformation controlled by a generated vector field, as shown in Figure \ref{diagram}. Mathematically,
\begin{equation}
\phi = G(x_1), ~~~\hat{x}_{0}(p) = x_1(p + \phi(p)), ~p\in X,
\end{equation}
\noindent where $G$ is a parameterized generator mapping from $x_1$ to vector field $\phi:X\rightarrow X$. Since $p + \phi(p)$ can lie between the set of coordinates on the grid for which we have defined values in $x_1$, we use bilinear interpolation when $X\in \mathbb{R}^2$, and trilinear interpolation when $X\in \mathbb{R}^3$. When $p + \phi(p)$ lies outside the range for which values are defined in $x_1$, values in $\hat{x}_{0}$ are set to $0$.

% adversarial training

To learn the parameters of transformation $f$, we follow~\cite{vagan} and use adversarial training. We train $G$ jointly with a parameterized critic $D$ tasked with discriminating modified images $\hat{x}_0$ from real images $x_0$. The generator $G$ is trained to fool $D$, using the gradient signal from $D$ so that the distribution of $\hat{x}_0$ approaches the distribution of $x_0$. For the adversarial loss, we use the \gls{WGAN}~\cite{wgan} formulation. The critic $D$ is trained to give high scores for modified images and low scores for real images, resulting in
\begin{equation}
L_D = L_{Dx_0} - L_{D\hat{x}_0}= \mathbb{E}\left[D(x_0)\right] -  \mathbb{E}\left[D(\hat{x}_{0})\right].
\end{equation}
\noindent To enforce the Lipschitz constraint required by the \gls{WGAN} formulation~\cite{wgan}, we use the penalty proposed in~\cite{improvedwgan},
\begin{equation}
L_{RegD} =\mathbb{E}\left[(\left\lVert \nabla_{\tilde{x}} D(\tilde{x})  \right\rVert_2 - 1)^2 \mid \tilde{x} \sim (1-\alpha)x_0 + \alpha \hat{x}_{0}, \alpha\sim\mathcal{U}(0,1)\right],
\end{equation}
\noindent where $\mathcal{U}(0,1)$ is the uniform distribution with support between $0$ and $1$. The expectation is approximated by sampling one $\alpha$ for each sampled $(x_0,x_1)$ pair. 

The generator $G$ is trained to lower the scores for $\hat{x}_0$, resulting in
\begin{equation}
L_G = \mathbb{E}\left[D(\hat{x}_{0})\right].
\end{equation}
\noindent To penalize extraneous changes and enforce smooth and realistic deformations, we use a total variation denoising~\cite{tv} penalty over $\phi$. This term forces nearby vectors to be aligned by penalizing $\int_{X}{\left\lVert \nabla \phi \right\rVert_2}$. After discretization,
\begin{equation}
L_{RegG} = \mathbb{E}\left[\frac{1}{P}\sum\nolimits_{p\in X}{\sum\nolimits_{n\in\mathcal{N}(p)}{\left\lVert \phi(p)-\phi(n)\right\rVert_2}}\right],
\end{equation}
\noindent where $\mathcal{N}(p)$ is the set of neighboring pixels of pixel $p$ (4-neighborhood for $X\in \mathbb{R}^2$ and 6-neighborhood for $X\in \mathbb{R}^3$) and $P$ is the total number of pixels.

The complete optimization formulation is given by
\begin{equation}
G^* = \argmin_{G} (L_{G}+\lambda_{RegG} L_{RegG} ), D^* = \argmin_{D} (L_{D}+\lambda_{RegD} L_{RegD}),
\end{equation}
where $\lambda_{RegG}$ and $\lambda_{RegD}$ are hyperparameters. 
\section{Experiments}
We tested our method\footnote{The \gls{DeFI} code is available at \url{https://github.com/ricbl/defigan}} in two datasets, one for finding evidence of \gls{AD} in brain MRIs and one for detecting signs of \gls{COPD} in \glspl{CXR}. We used as a baseline the \gls{VA-GAN} method~\cite{vagan}, where the function to obtain $\hat{x}_0$ is defined by the addition of a difference map, i.e., $\hat{x}_{0} = x_1+G(x_1)$, and the penalty on $G$ is accordingly defined by $L_{RegG} = \left\lVert \hat{x}_{0} - x_1 \right\rVert_{1}$.

We used u-nets~\cite{unet} as $G$'s architecture, varying the number of downsamplings and channels depending on the dataset. We used the Adam optimizer~\cite{adam} with a learning rate of $1\mathrm{e}{-4}$  and set $\lambda_{RegD}=10$. For $D$'s architecture, we used an ImageNet pre-trained Resnet-18~\cite{resnet} for the \gls{CXR} dataset and a 10-layer network as in~\cite{vagan} for the MRI dataset. For training the \gls{DeFI} method, we performed 100 updates to $D$ for each update to $G$, whereas for \gls{VA-GAN}, the 100:1 update ratio was used for the first 25 updates to $G$, and then changed to a 5:1 ratio. It was important for the convergence of $G$ to have $X$ represented in pixel units.
 
\subsection{Chronic obstructive pulmonary disease in chest x-rays}

We collected a dataset containing \gls{PA} \glspl{CXR} acquired at the University of Utah Hospital from 2012 to 2017. Each \gls{CXR} was labeled for COPD using results from \glspl{PFT} taken within one month of the \gls{CXR}. Patients who received a lung transplant were excluded. Patients with \gls{COPD} were assigned to class 1, and others were assigned to class 0. The training set was composed of patients who had only one \gls{CXR} included in the dataset and contained 2,226 images from normal patients and 963 images from \gls{COPD} patients. We worked under an approved \gls{IRB}\footnote{IRB\_00104019, PI: Schroeder MD} and anonymized data using Orthanc\footnote{\url{orthanc-server.com}}. Image preprocessing included center-cropping to a square, resizing to 256$\times$256, cropping to 224$\times$224 (randomly for training and centered for validation and testing), and equalizing histograms. 

\subsection{Alzheimer's disease in brain MRIs}

The \gls{ADNI} dataset was collected to characterize the progression of \gls{AD} and contains brain MRIs of thousands of subjects followed for a few years~\cite{adni}. Diagnosis of \gls{AD} and \gls{MCI} are provided for each exam. Following~\cite{vagan}, we set class 0 to \gls{MCI} patients, instead of healthy patients, and class 1 to \gls{AD} patients, training $G$ to provide interpretations for the conversion from \gls{MCI} to \gls{AD}. We also limited our study to T1-weighted MRIs from the ADNI1, ADNIGO, and ADNI2 cohorts.

Data splits and part of the preprocessing were replicated from~\cite{vagan}\footnote{\gls{VA-GAN}'s original preprocessing code, list of subjects, and TensorFlow implementation are available in \url{https://github.com/baumgach/vagan-code/}}. For each image, we used the \gls{fsl} toolbox~\cite{fsl} v6.0 for reorientation and \gls{fov} cropping, N4 Bias Correction~\cite{N4} from \gls{ants} v3.0 for bias field correction, \gls{fsl} to register the image to the MNI 152 space~\cite{mni2,mni1}, and ROBEX~\cite{robex} v1.2 for skull stripping. To correct cases where ROBEX failed, we added another step of skull stripping to the whole dataset using \gls{bet}~\cite{bet} with a fractional intensity threshold equal to $0.25$. All volumes were rescaled to a size of $128\times 160\times 112$ pixels. Through visual inspection, we searched for cases where the brain volume was incorrectly oriented, or skull stripping had failed, and removed 7 volumes from 4 subjects from test and validation. The preprocessing differed from \cite{vagan} in the application of \gls{bet} and in the elimination of the 7 cases for which preprocessing failed. The final training set had 2,528 \gls{MCI} volumes and 1,198 \gls{AD} volumes, for a total of 825 subjects.

\subsection{Quantitative Validation}

We validated our generated disease evidence by using longitudinal scans demonstrating disease progression. For the \gls{COPD} dataset, we paired all test cases where a subject was healthy in the baseline \gls{CXR} with cases of the same subject after developing \gls{COPD}. For each pair, we performed an affine registration of the normal case to the \gls{COPD} case using SimpleITK v1.2~\cite{sitk}. We then subtracted the \gls{COPD} image from the registered baseline image to get a difference ground truth $\mathsmaller{\Delta} x$. The final split sizes were 206 pairs of images (176 subjects) for validation and 547 pairs (354 subjects) for testing. For the \gls{ADNI} dataset, we performed rigid registration instead of affine, disregarding the background for the calculations, and used \gls{MCI} cases as the baseline, pairing them with \gls{AD} MRIs taken with the same field strength. The final split sizes were 207 pairs (102 subjects) for validation and 259 pairs (143 subjects) for testing.

The ground truth $\mathsmaller{\Delta} x$ and the predicted difference $\widehat{\mathsmaller{\Delta} x}=\hat{x}_{0} - x_1$ were compared using \gls{ncc}, masked to ignore regions pad\-ded during registration and skull stripping. This operation was defined by
\begin{equation}
\text{NCC}\left( \textstyle\mathsmaller{\Delta} x, \widehat{ \mathsmaller{\Delta} x}\right)=\frac{1}{M-1}\sum_{p\in \mathcal{M} }{\frac{\left({\mathsmaller{\Delta} x}\left(p\right)-\mu_{\mathsmaller{\Delta} x}\right)}{\sigma_{\mathsmaller{\Delta} x}}\times\frac{\left(\widehat{\mathsmaller{\Delta} x}\left(p\right)-\mu_{\widehat{\mathsmaller{\Delta} x}}\right)}{ \sigma_{\widehat{\mathsmaller{\Delta} x}}}},
\end{equation}
\noindent where $M$ is the number of pixels in the mask $\mathcal{M}$, and $\mu_i$ and $\sigma_i$ are, respectively, the average value and the unbiased standard deviation of values of pixels of image $i$ inside mask $\mathcal{M}$. One subject may have had more than one pair of validation images. The score calculated for all pairs of each subject was averaged before calculating the final average score to avoid the over-representation of a few subjects in the final score.

Hyperparameters were chosen by considering image quality and validation \gls{ncc} score. With the \gls{DeFI} method, we used $\lambda_{RegG}=25$ for the \gls{COPD} dataset and $\lambda_{RegG}=10$ for the \gls{ADNI} dataset, whereas for the \gls{VA-GAN} method we used, respectively, $\lambda_{RegG}=50$  and $\lambda_{RegG}=100$. The original \gls{VA-GAN} TensorFlow implementation was used for the \gls{ADNI} dataset. %This implementation had the peculiarity of using $\hat{x}_{0} = \tanh(x_1+G(x_1))$, in place of Equation \ref{vaganaddition}, and a learning rate of $1\mathrm{e}{-3}$.

\begin{table}[t]  
\centering
\caption{\gls{ncc} scores for the compared methods. Averages of 5 models trained with different random seeds are reported with their standard deviations.}
\begin{tabular}{@{}lcc@{}}
\toprule
\bsfrac{Method}{Dataset} & \gls{VA-GAN}   & \gls{DeFI}           \\ \midrule
\gls{ADNI}                          & $0.332\pm 0.015$ & $\bm{0.365\pm 0.014}$ \\
\gls{COPD}                          & $0.174\pm 0.007$ & $\bm{0.204\pm 0.006}$ \\ \bottomrule
\label{tableresults}
\end{tabular}
\end{table}
Test scores are reported in Table \ref{tableresults}, where the \gls{DeFI} method is shown to outperform the baseline. The scores for the \gls{ADNI} dataset when using \gls{VA-GAN} are better than the ones presented in \cite{vagan}. This difference is probably due to choosing the best training epoch using the \gls{ncc} validation score instead of loss values. It is interesting to note that training was less noisy for the \gls{DeFI} method when considering the validation scores from epoch to epoch. Relatively low \gls{ncc} scores may result from the difficulty in aligning longitudinal data and from aging effects in the ground truth.

\subsection{Biomarker Validation}

 \begin{figure}[t]
 \centering
\includegraphics[width=1.0\textwidth]{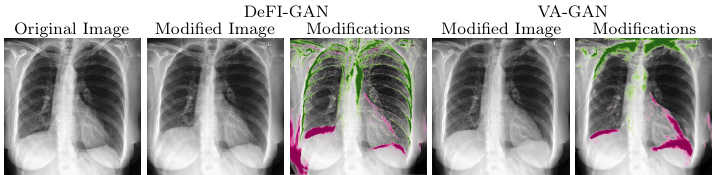}
\caption{Comparison of results of \gls{DeFI} and \gls{VA-GAN} on the \gls{COPD} dataset. Green represents darkening of the image, and pink represents brightening.} \label{macroxray}
\end{figure}

 \begin{figure}[t]
 \centering
\includegraphics[width=1.0\textwidth]{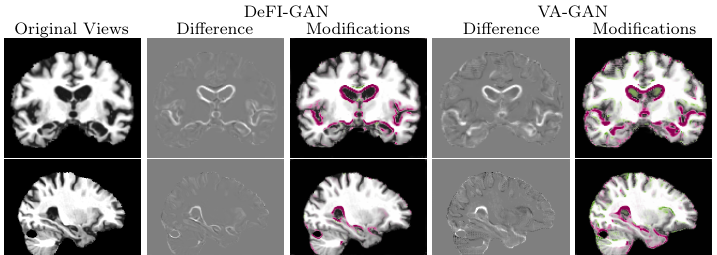}
\caption{Results on the \gls{ADNI} dataset when removing evidence of \gls{AD} from the original image. Green represents darkening of the image, and pink represents brightening.} \label{macrobrain}
\end{figure}

We compared the interpretations provided by \gls{DeFI} and \gls{VA-GAN} and present some results in Figures \ref{macroxray} and \ref{macrobrain}. The difference maps were computed using $\hat{x}_{0} - x_1$. Additional results are presented in the supplementary material. For the \gls{COPD} dataset, both approaches identified changes in heart silhouette and in diaphragm height and curvature, which are notable characterizations of \gls{COPD} in \glspl{CXR}~\cite{emphysema1,emphysema2}. We found no variation in lung lucency. The lack of change in texture was expected for the \gls{DeFI} method from its formulation. Interpretations from the \gls{VA-GAN} method were less spatially smooth.

 \begin{figure}[t]
 \centering
\includegraphics[width=1.0\textwidth]{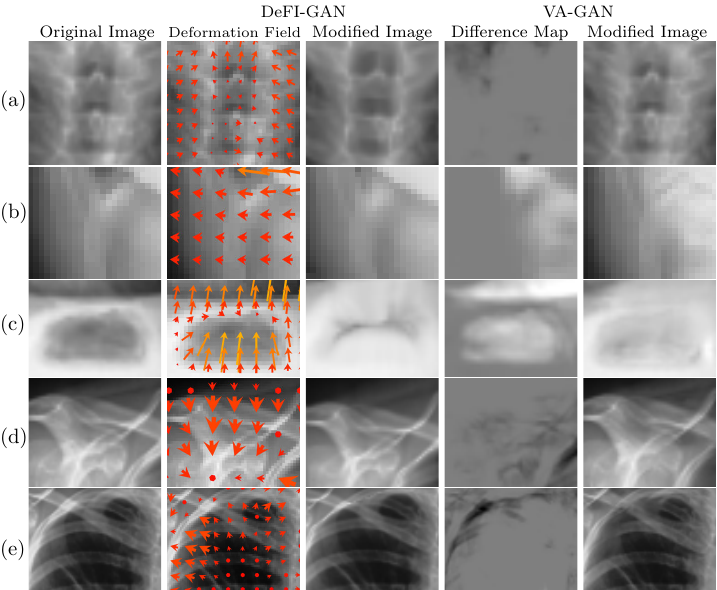}

\caption{Details on the generated deformation fields and comparison with \gls{VA-GAN}. The color of the arrows represents their length. For rows \textbf{a-c}, their length represents the exact shift in pixels. In rows \textbf{d-e}, their length is multiplied by 2 for better visualization. The full-size images are presented in the supplementary material. From top to bottom: \textbf{(a)} trachea; \textbf{(b)} soft tissue; \textbf{(c)} gastric bubble; \textbf{(d)} shoulder; \textbf{(e)} lung apex.} \label{details}
\end{figure}

Saber-sheath trachea is a secondary sign of \gls{COPD}, and, when it occurs, there is a narrowing of the trachea in the frontal \gls{CXR}~\cite{saber}. This evidence was more identifiable in models trained with \gls{DeFI} when looking at quiver plot visualization, as seen in Figure \ref{details}(a). The \gls{VA-GAN} method cannot produce this kind of deformation field visualization. In the difference map produced by the \gls{VA-GAN} method, the changes in the top of the image appeared as a dark fuzzy alteration. 
Another sign more easily identified with \gls{DeFI} was the change in soft-tissue thickness, as shown in Figure~\ref{details}(b). This change is in accordance with the correlation between low \gls{BMI} and \gls{COPD}~\cite{bmi}.

Although we focused the analysis on \gls{COPD}, we also found the results in the \gls{ADNI} dataset consistent with the literature. The main changes associated with \gls{AD}, expansion of ventricles, hippocampus atrophy, and temporal lobe atrophy~\cite{brain_main}, are seen in Figure \ref{macrobrain}.  It was possible to find, in some cases, atrophies in the precuneus, the cerebellum, and the brainstem. These regions have been associated with \gls{AD}~\cite{precuneus,cerebellum,brainstem}. The rest of the cortex also presented atrophies to a lesser extent. The differences highlighted by both methods were similar, except for \gls{VA-GAN} having more image noise, more darkening differences, and more noticeable highlighting of the hippocampus, which is favorable evidence. A detailed representation of a deformation field in MRIs can be seen in the supplementary material.
\subsubsection{Unexpected changes in \gls{COPD}} Some evidence pinpointed by the \gls{DeFI} method in the \gls{COPD} dataset is not usually listed in the literature. Figure~\ref{details}(c) shows a decrease in gastric bubble size. This change may represent a confusion between the lower border of the gastric bubbles and the diaphragm, or be related to aerophagia due to coughing~\cite{aerophagia}. Furthermore, the shoulder position changed in a few images, as seen in Figure \ref{details}(d). High clavicle position might be related to a decrease in shoulder mobility~\cite{shoulder} and, consequently, difficulty in following the required body positioning for the standing \gls{CXR} acquisition. It may also be related to a bias in spine position due to the prevalence of osteoporosis~\cite{osteo} in patients with \gls{COPD}. In Figure \ref{details}(e), the expansion of the top of the lungs contradicts larger lung volumes in \gls{COPD} patients. However, when analyzed by a radiologist, lung volume still seemed larger in the original images, probably due to diaphragm height. This change may be correlated with the clavicle positioning since it increases lung volume above the clavicle. The change in gastric bubble size and the expansion of the top of the lung were, in a less comprehensible representation, also highlighted by the \gls{VA-GAN} method. All these unexpected differences may also represent dataset biases. A more in-depth investigation of the causes for these changes is left to subsequent studies. 

\section{Conclusion}
We formulated a method for producing interpretations of the disease evidence that deep learning models can capture from imaging data, using deformation fields and adversarial training. This method performed better in the proposed longitudinal quantitative validation on two medical datasets, when compared to another method with a similar goal. Furthermore, it allowed for easier discovery of qualitative disease evidence used by the model. As exemplified by some of the highlighted \gls{COPD} evidence, this study has the potential to support future analyses of unexpected biases in medical imaging datasets.

\end{document}

\section*{Supplementary Material}
 \begin{figure}[H]
 \centering
\includegraphics[angle=90,width=0.68\textwidth,trim={1.2cm 0.5cm 1.2cm 0.5cm},clip]{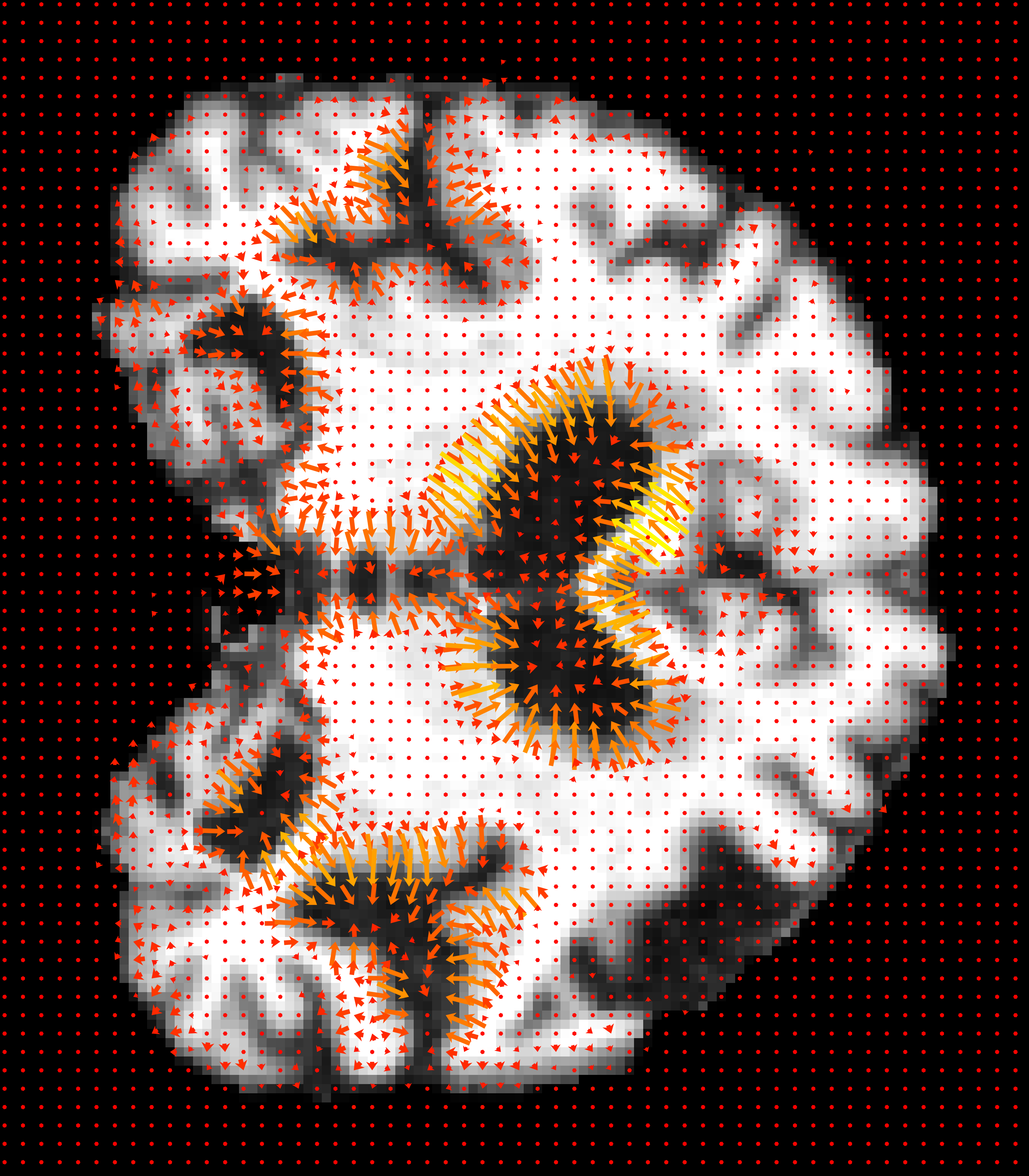}
\caption{Deformation field for an ADNI image. Arrow length is multiplied by 5.} \label{brainflow}
\end{figure}

 \begin{figure}[H]
 \centering
\includegraphics[width=1.0\textwidth]{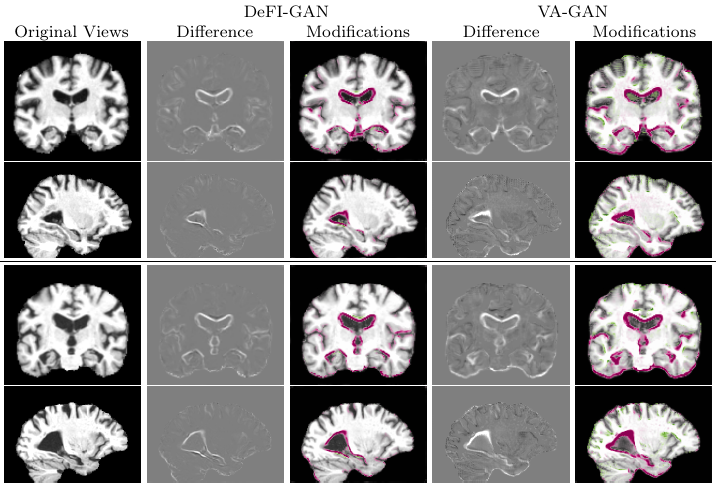}
\caption{Additional results for the  ADNI dataset.} \label{morebrain}
\end{figure}

 \begin{figure}[H]
 \centering
 \textbf{A)}
\adjincludegraphics[width=0.28\textwidth, valign=t]{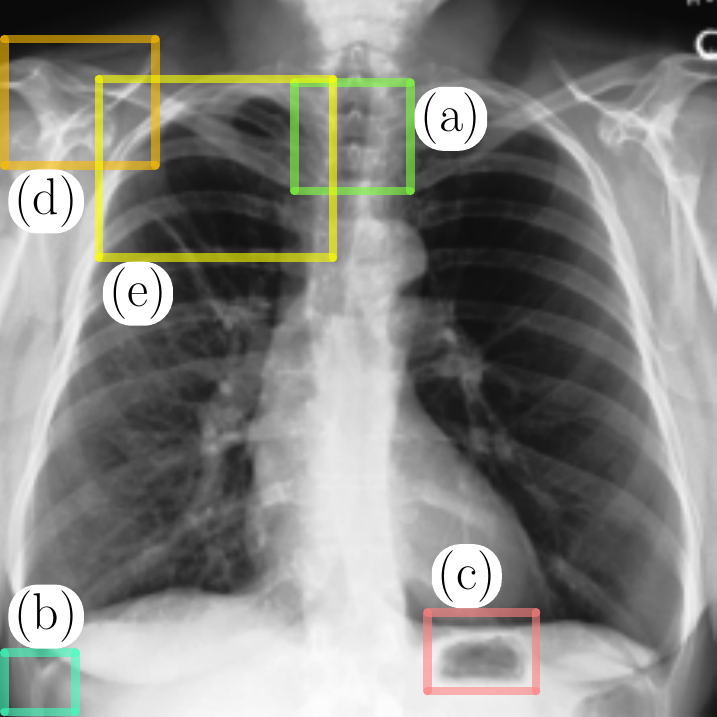}
\textbf{B)}
\adjincludegraphics[width=0.28\textwidth, valign=t]{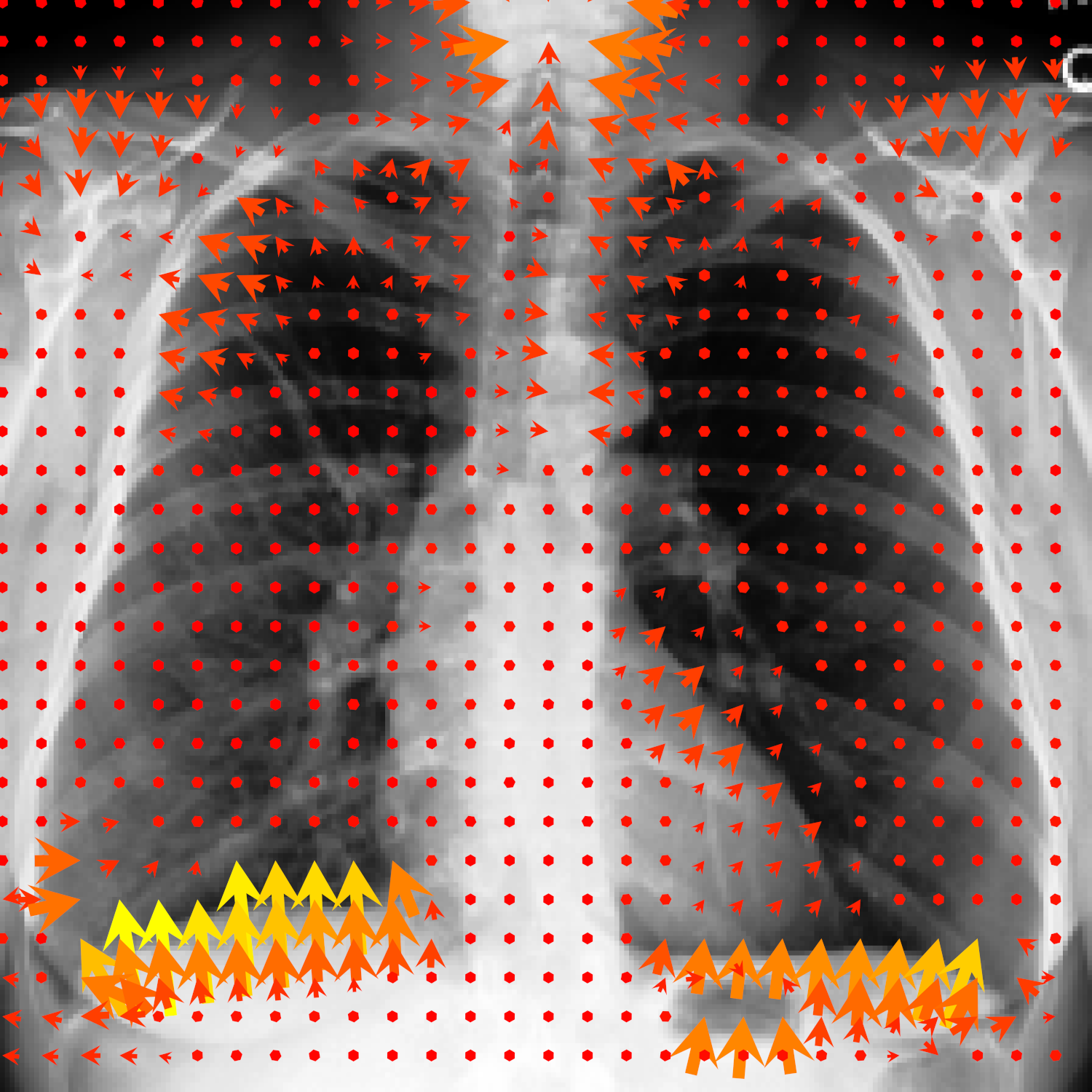}
\textbf{C)}
\adjincludegraphics[width=0.28\textwidth, valign=t]{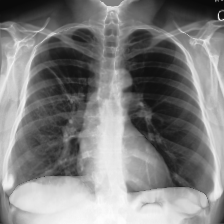}
\textbf{D)}
\adjincludegraphics[width=0.28\textwidth, valign=t]{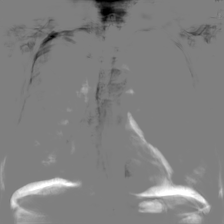}
\textbf{E)}
\adjincludegraphics[width=0.28\textwidth, valign=t]{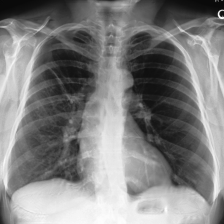}
\caption{\label{detailsplus}Full-size images of the portions used in Figure 4 in the main paper: \textbf{A)} cropped regions labeled by their row labels over the original image; \textbf{B)} deformation field with arrow length scaled by 2; \textbf{C)} image modified by \gls{DeFI}; \textbf{D)} \gls{VA-GAN} difference map; \textbf{E)} image modified by \gls{VA-GAN}.}
\end{figure}

 \begin{figure}[H]
 \centering
\includegraphics[width=1.0\textwidth]{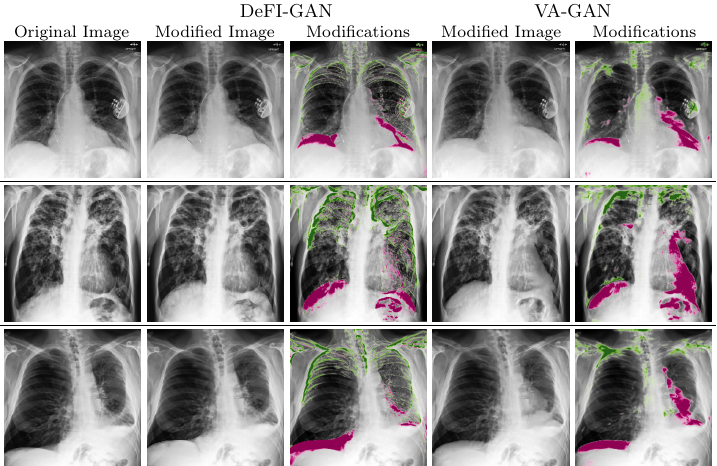}
 \centering
\caption{Additional results for the COPD dataset.} \label{morexray}
\end{figure}